\documentclass[twocolumn,pra,aps,superscriptaddress,showpacs]{revtex4}

\usepackage[T1]{fontenc}
\usepackage{calc}
\usepackage{mathtools}
\usepackage{bm}
\usepackage{amsmath}
\usepackage{amssymb}
\usepackage{stackrel}
\usepackage{graphicx}
\usepackage{setspace}
\usepackage{subfigure}
\usepackage{indentfirst}
\usepackage{cases}
\makeatletter

\newcommand{\Rmnum}[1]{\expandafter\@slowromancap\romannumeral #1@}

\makeatother

\begin{document}

\title{Quantum temporal steering in a dephasing channel with quantum criticality}

\author{B. Liu}

\affiliation{Department of Physics, Hangzhou Normal University, Hangzhou, 311121, China}

\affiliation{School of Physics, Changchun Normal University, Changchun, 130032, China}

\author{Y.\,X. Huang}

\affiliation{School of Science, Zhejiang University of Science and Technology, Hangzhou, 310023, China}

\author{Z. Sun}

\email{sunzhe@hznu.edu.cn}

\affiliation{Department of Physics, Hangzhou Normal University, Hangzhou, 311121, China}

\begin{abstract}
  We investigate the quantum temporal steering (TS), i.e., a temporal analogue
of Einstein-Podolsky-Rosen steering, in a dephasing channel which is modeled
by a central spin half surrounded by a spin-1/2 \textit{XY} chain where
quantum phase transition happens. The TS parameter $S_{\text{TS}}$ and the
TS weight $W_{\text{TS}}$ are employed to characterize the TS dynamics. We
analytically obtain the dependence of $S_{\text{TS}}$ on the decoherence
factor. The numerical results show an obvious suppression of $S_{\text{TS}}$
and $W_{\text{TS}}$ when the \textit{XY} chain approaches to the critical
point. In view of the significance of quantum channel, we develop a new
concept, \textit{TS weight power}, in order to quantify the capacity of the
quantum channel in dominating TS behavior. This new quantity enables us to
indicate the quantum criticality of the environment by the quantum
correlation of TS in the coupled system.
\end{abstract}
\maketitle

\section{Introduction}
Recently, the problems of quantum steering have attracted considerable
interest theoretically and experimentally~\cite%
{1,2,5,3,4,6,7,8,QYHe1,9,10,QYHe2,11,12,14,15,16,17,18,19}. In a bipartite
system sharing entangled states, Einstein-Podolsky-Rosen (EPR) steering
problems refer to the quantum nonlocal correlations which allows one of the
subsystems to remotely prepare or steer the other one via local
measurements. Apart from the fundamental significance of steering in quantum
mechanics~\cite{1,2,5,3,4,6,7,8}, there are lots of application motivations
that EPR steering was thought to be the driving power of quantum
cryptography and teleportation~\cite{9,10,QYHe2,11}. For example, steering
allows quantum key distribution (QKD) if one party trusts his own devices
but not that of the other party~\cite{11}. An obvious advantage of this
steering-dominated scenario is easier to implement, when compared with the
device-independent protocols~\cite{12}.

In order to detect steerability, several inequalities in terms of the
sufficient conditions of steerability were employed~\cite{1,2,5,3}, and such
inequalities have been tested by several experiments~\cite{3,4,9}. Beyond
inequalities, a range of possible measures were developed to quantify the
amount of steering~\cite{14}, then the existence of steerability can be
sufficiently and necessarily demonstrated. Very recently, a powerful measure named the steerable weight was proposed~\cite%
{15,16}. In other respects, steerability was found to be equivalent to joint
measurability~\cite{17,18}. A close relationship between steerability and
quantum-subchannel discrimination was discovered in~\cite{19}.

Instead of discussing spatially-separated systems, a novel and important
direction is to consider the quantum steering in time, i.e., the so-called
temporal steering (TS)~\cite{20}, which focuses on a single object at
different times. In this frame, a system is sent to a distant receiver (say
Bob) through a quantum channel, then a detector or manipulator (say Alice)
can perform some operations including measurements on the system, before Bob
receives the system and performs his measurement. The existence of nonzero
TS indicates that the temporal correlations, accounting for how strong
Alice's choice of measurements at an initial time, can influence the final
state captured by Bob. This kind of correlation can be detected by the TS
inequality\thinspace \cite{20,20-1,20-2}, which also becomes a very useful
tool in verifying the suitability of a quantum channel for a certain QKD
process. Compared to the spatial steering inequalities, TS inequality
presents a superior applicability that the entanglement-based scenarios are
no longer necessary, and one can directly carry out the BB84 protocol and
other related protocols.

In order to quantify TS, a concept of TS weight was introduced in the literature~\cite{21}, where the authors found that TS
characterized by the weight can be used to define a sufficient and practical
measure of strong non-Markovianity. Therefore, it also implies the evident
dependence of TS dynamics on the properties of quantum channels. Another temporal steering quantify called temporal steering robustness was introduced in by Ku \emph{et al.}~\cite{Ku16}, where TS was used for testing magnetoreception. The TS problems were also studied experimentally in a phenomenologically designed channel~\cite{22} and in a proposed channel based on the spin-boson interaction without rotating-wave rotating approximations~\cite{Xiong17}. In other aspects, TS was found intrinsically associated with realism and joint measurability~\cite{23,24}. The hierarchy among the temporal inseparability, temporal steering, and macrorealism was found~\cite{Ku17}. Moreover, there is a recent work which studies the spatial-temporal steering~\cite{Chen17,Costa17} and applied for testing quantum correlations in quantum networks including biological nano-structures.

In this paper, we intensively investigate the influence of the quantum
channel on the TS behavior. This is a particular problem in TS other than
EPR steering, because the formation of the quantum correlation between the
system's initial and final state lies on the quantum-channel. Therefore, TS
is sensitive to the characteristic of the channel, then an interesting
question arises: How does the TS behave in a special environment where
quantum phase transition (QPT) happens?

It is known that QPT takes place at zero temperature, at which the thermal
fluctuations vanish. Therefore, quantum fluctuation plays the major role in
QPT. At the critical point, a qualitative change occurs in the ground state
and long-range correlation also develops. Actually, if QPTs appear in the
surrounding, the coherence properties of the center system indeed suffer
significance changes, e.g., it enhances the decoherence of the system~\cite
{quan,Zhang,XSMa,BShao,Yuan}, accelerates the disentanglement~\cite
{Zhe07} and also helps to induce quantum correlations~\cite{X.X.Yi,QAi}.
Motivated by these, we will take into account the influence of the quantum
criticality on the dynamics of quantum TS.

\section{Two-dimensional dephasing channel}
Let us start from a general dephasing model. For a two-dimensional system,
the two basis vectors $\left\vert 1\right\rangle $ and $\left\vert
0\right\rangle $ are chosen as the eigenstates of the Pauli operator $\sigma
_{z}$ obeying $\sigma _{z}\left\vert 1\right\rangle =\left\vert
1\right\rangle $\ and $\sigma _{z}\left\vert 0\right\rangle =-\left\vert
0\right\rangle $. The interaction Hamiltonian reads%
\begin{equation}
H=\left\vert 1\right\rangle \left\langle 1\right\vert \otimes
H_{+}+\left\vert 0\right\rangle \left\langle 0\right\vert \otimes H_{-},
\label{H_1}
\end{equation}%
where $H_{\pm }$ denotes the resulting Hamiltonian corresponding to the
eigenvalues of the qubit system. Thus the time evolution operator becomes%
\begin{equation}
U(t)=\left\vert 1\right\rangle \left\langle 1\right\vert \otimes
U_{+}(t)+\left\vert 0\right\rangle \left\langle 0\right\vert \otimes
U_{-}(t).  \label{U_tot}
\end{equation}%
The initial state of the total system is assumed in a product form%
\begin{equation}
\rho _{\text{tot}}(0)=\rho _{s}\left( 0\right) \otimes \rho _{e}\left(
0\right) ,
\end{equation}%
where the states $\rho _{s}$ and $\rho _{e}$ denote the system and
environment state respectively. Under the time evolution defined in
Eq.\thinspace (\ref{U_tot}) we finally obtain the reduced density matrix of
the system
\begin{equation}
\rho _{s}\left( t\right) =\left(
\begin{array}{cc}
\rho _{11}\left( 0\right)  & \rho _{12}\left( 0\right) F_{t}^{\ast } \\
\rho _{21}\left( 0\right) F_{t} & 1-\rho _{11}\left( 0\right)
\end{array}%
\right) ,  \label{rho_t}
\end{equation}%
where $\rho _{ij}\left( 0\right) $ ($i,j=1,2$) denote the elements of the
initial state of the system.\ The decoherence factor, defined by $F\left(
t\right) =$Tr$\left[ \rho _{e}\left( 0\right) U_{+}^{\dagger }(t)U_{-}(t)%
\right] $, plays a key role in the dynamic process.

\section{System-bath Hamiltonian with quantum criticality}
A spin-1/2 \textit{XY} spin chain displays the surrounding system where rich
QPTs happen, and the Hamiltonian reads

\begin{equation}
H_{e}=-\sum_{l=-M}^{M}\left( \frac{1+\gamma }{2}\sigma _{l}^{x}\sigma
_{l+1}^{x}+\frac{1-\gamma }{2}\sigma _{l}^{y}\sigma _{l+1}^{y}+{\lambda }%
\sigma _{l}^{z}\right) ,  \label{hhh}
\end{equation}%
where ${\lambda }$ characterizes the strength of the transverse field, and $%
\gamma $ is the dimensionless anisotropy parameter. $\sigma _{l}^{\alpha
}\left( \alpha =x,y,z\right) $ are the Pauli operators defined on the $l$-th
site, and the total number of spins in the chain is $L=2M+1$. The periodic
boundary condition $\sigma _{L+1}=\sigma _{1}$ is assumed here and we
henceforth set $\hbar =1$.\

This \textit{XY} model becomes the transverse Ising chain for $\gamma =1$.
The critical point is at $\left\vert \lambda _{c}\right\vert =1$. Before
that, when $\left\vert \lambda \right\vert \ll 1$, there are doubly
degenerate ground states with all spins pointing either up or down along the
$x$ (or $y$) axis. Whereas, for large cases of $\left\vert \lambda
\right\vert \gg 1$, the spins of ground state are polarized up along $z$
direction. Therefore, at the critical point $\left\vert \lambda
_{c}\right\vert =1$, there is a fundamental change of the ground state. When
the parameter $\gamma =0$, it is the \textit{XX} model. There is a
criticality region along the line between $\lambda =-1$ and $\lambda =1$.

If one characterizes the critical features in terms of a critical exponent $%
\nu $ defined by the correlation length $\delta =\left\vert \lambda -\lambda
_{c}\right\vert ^{-\nu }$. For any values of $\gamma $, quantum criticality
occurs at the critical field $\lambda _{c}=1$. For the interval $0<\gamma
\leq 1$ the model belongs to the Ising universality class corresponding to
the critical exponent $\nu =1$, while for $\gamma =0$ the model belongs to
the \textit{XX} universality class with $\nu =1/2$~\cite{S.Sachdev}.

Fortunately, this \textit{XY} model can be calculated exactly. By combining
Jordan-Wigner transformation via $\prod\nolimits_{k<l}\sigma _{k}^{z}\sigma
_{l}^{+}=c_{l}$ and Fourier transformation via $d_{k}=\frac{1}{\sqrt{N}}%
\sum_{l=-M}^{M}e^{-i\frac{2\pi lk}{N}}c_{l}$\ to the momentum space~\cite%
{S.Sachdev}, the Hamiltonian can be written in a diagonal form as~\cite%
{YDWang,Zhe07}
\begin{equation}
H_{e}=\sum_{k>0}\left( \Omega _{k}e^{i\frac{\theta _{k}}{2}\tilde{\sigma}%
_{kx}}\sigma _{kz}e^{-i\frac{\theta _{k}}{2}\tilde{\sigma}_{kx}}\right) +({%
\lambda -1)}\tilde{\sigma}_{0z},  \label{diag_H}
\end{equation}%
where we have used the following pseudospin operators $\tilde{\sigma}%
_{k\alpha }\left( \alpha =x,y,z\right)$~\cite{Anderson}
\begin{eqnarray}
\tilde{\sigma}_{kx} &=&d_{k}^{\dagger }d_{-k}^{\dagger }+d_{-k}d_{k},\left(
k=1,2,...M\right) \   \notag \\
\tilde{\sigma}_{ky} &=&-id_{k}^{\dagger }d_{-k}^{\dagger }+id_{-k}d_{k},
\notag \\
\tilde{\sigma}_{kz} &=&d_{k}^{\dagger }d_{k}+d_{-k}^{\dagger }d_{-k}-1,
\notag \\
\tilde{\sigma}_{0z} &=&2d_{0}^{\dagger }d_{0}-1,
\end{eqnarray}%
and $d_{k}^{\dagger },d_{k}\{k=0,1,2,...\}$\ denote the fermionic creation
and annihilation operators in the momentum space, respectively. Here,

\begin{eqnarray}
\Omega _{k} &=&-2\sqrt{[{\lambda }-\cos (2\pi k/L)]^{2}+\gamma ^{2}\sin
^{2}(2\pi k/L)},  \notag \\
\theta _{k} &=&\arcsin \left[ \frac{2\gamma \sin \left( 2\pi k/L\right) }{%
\Omega _{k}}\right] .  \label{p2}
\end{eqnarray}%
When an auxiliary qubit system is transversely coupled to the chain through
the interaction Hamiltonian described by
\begin{equation}
H_{I}={g\sigma }_{z}\sum_{l=-M}^{M}\sigma _{l}^{z},
\end{equation}%
where operator ${\sigma }_{z}$ displays the center qubit system, and $g$
denotes the coupling strength. Then the total Hamiltonian becomes $%
H=H_{e}+H_{I}$, where we safely ignore the free Hamiltonian of the auxiliary
particle. The $H=\left\vert 1\right\rangle \left\langle 1\right\vert \otimes
H_{+}+\left\vert 0\right\rangle \left\langle 0\right\vert \otimes H_{-}$ in
Eq.\thinspace (\ref{H_1}) can be constructed, where $H_{\pm }$ can be
obtained by replacing ${\lambda }$ with ${\lambda }_{\pm }={\lambda \pm g}$
in Eq.\thinspace (\ref{diag_H}) and (\ref{p2}).

The evolution is described as:
\begin{equation}
U_{\pm }(t)=e^{i(1-{\lambda }_{\pm }{)}\tilde{\sigma}_{0z}t}\prod_{k>0}e^{i%
\frac{\theta _{k\pm }}{2}\tilde{\sigma}_{kx}}e^{-it\Omega _{k\pm }\tilde{%
\sigma}_{kz}}e^{-i\frac{\theta _{k\pm }}{2}\tilde{\sigma}_{kx}},
\end{equation}%
where the parameters $\Omega _{k,\pm }$ and $\theta _{k,\pm }$ can be
obtained by replacing ${\lambda }$ with ${\lambda }_{\pm }={\lambda \pm g}$
in Eq.\thinspace (\ref{p2}). When we focus on the dynamics of the center system by doing partial trace over the bath space, the decoherence factor $F_t$ ( $F^*_t$) appears at the non-diagonal element of the reduced density matrix of the system with the definition $F_t=\text{Tr}[\rho_{e}U^{\dagger}_{-}U_{+}]$, which characterizes the dynamics of the quantum coherence.
If we assume that the initial state of the
environment is the ground state of the \textit{XY} spin chain
\begin{equation}
|\psi \rangle _{e}=|1\rangle _{k=0}\otimes _{k>0}e^{i\frac{\theta _{k}}{2}%
\sigma _{kx}}|1\rangle _{k}|1\rangle _{-k},
\end{equation}%
where it should be noted that the first term $|1\rangle _{k=0}$ is for the
case $\lambda <1,$ while $\lambda \geq 1$, it should be $|0\rangle _{k=0}$.
Subsequently we obtain the decoherence factor
\begin{eqnarray}
F_{t} &=&e^{-2itg}\prod_{k>0}\{\cos \left( \Omega _{k,-}t\right) \cos \left(
\Omega _{k,+}t\right)  \notag \\
&&+\sin \left( \Omega _{k,-}t\right) \sin \left( \Omega _{k,+}t\right) \cos
\left( \theta _{k,+}-\theta _{k,-}\right)  \notag \\
&&+i\cos \left( \Omega _{k,-}t\right) \sin \left( \Omega _{k,+}t\right) \cos
\theta _{k,+}  \notag \\
&&-i\sin \left( \Omega _{k,-}t\right) \cos \left( \Omega _{k,+}t\right) \cos
\theta _{k,-}\},  \label{Ft}
\end{eqnarray}

By introducing a cutoff number $K_{c}$ at the thermodynamic limit $%
L\rightarrow \infty $ and the critical limit $\lambda \rightarrow 1$, the
norm of the factor approximately decays with time as\thinspace \cite{quan}
(see details in the methods):
\begin{equation}
\left\vert F\right\vert _{K_{c}}\approx e^{-R_{c}t^{2}},  \label{Ft-smallk}
\end{equation}%
with $R_{c}=16E\left( K_{c}\right) \gamma ^{2}g^{2}\left( 1-\lambda \right)
^{-2}$ and $E( K_{c}) =2\pi ^{2}K_{c}( K_{c}+1) (2K_{c}+1) /( 6L^{2})$. Even at some cases that the factor $%
E\left( K_{c}\right) \gamma ^{2}g^{2}\rightarrow 0$, the decay of the $%
\left\vert F\left( t\right) \right\vert $ is still possible due to the
vanishing denominator $\left( 1-\lambda \right) ^{2}$ near the critical
point $\lambda _{c}=1$. This heuristic analysis approximately predicts a
monotonic decoherence process near the critical point $\lambda _{c}=1$ for
any values of $\gamma $, while, beyond the critical region, different
dynamics will be found.
\begin{figure}[tbp]
\begin{center}
\includegraphics[width=0.5\textwidth, height=0.4\textwidth]{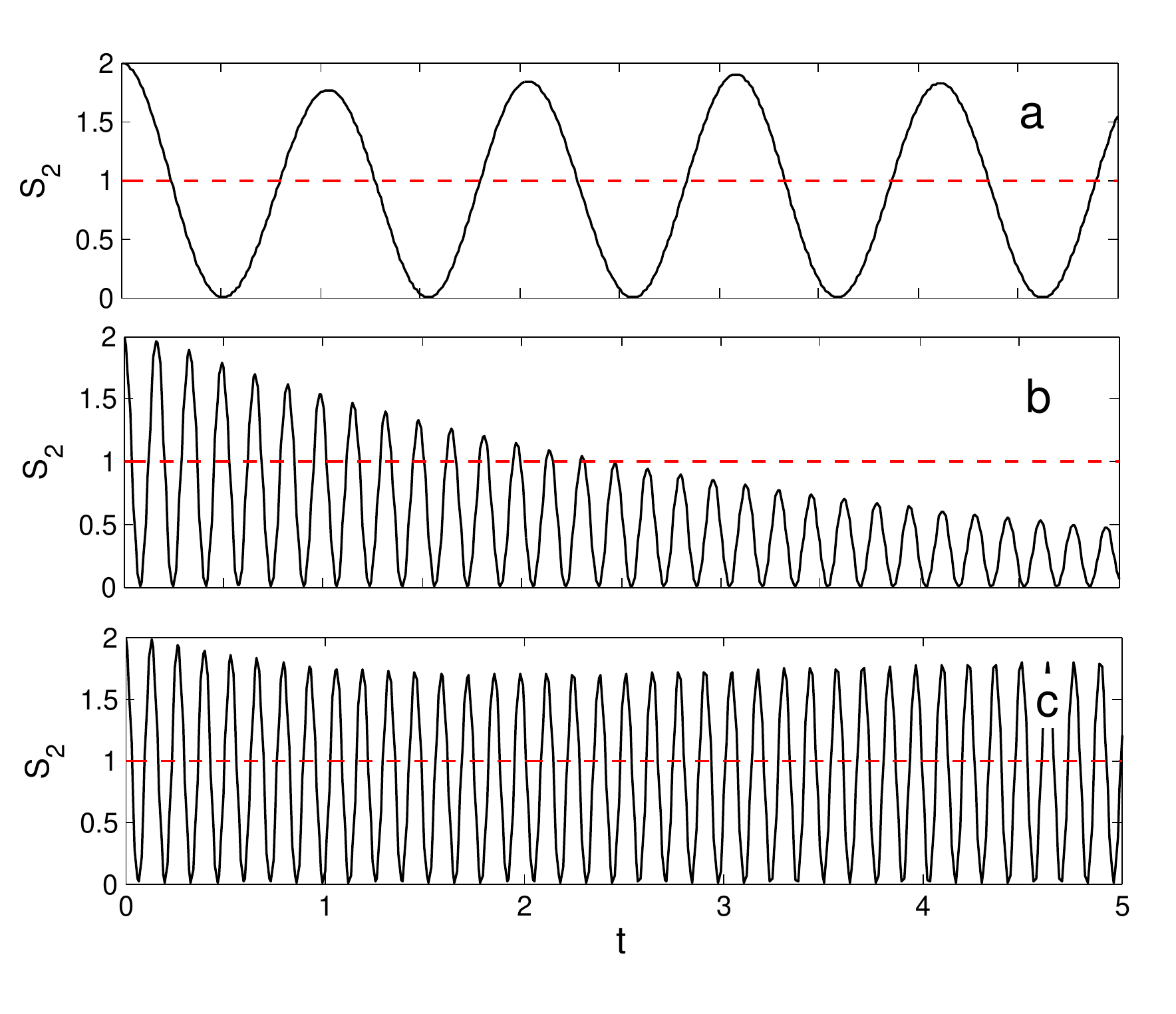}
\end{center}
\caption{
(Color online) Bidirectional-measurement TS parameter versus time $%
t $ for different parameters $\protect\lambda =0.2,~1$ and $1.2$ presented
in subfigures (a), (b) and (c), respectively. The parameter $\protect\gamma %
=1$, the size of the surrounding spin chain is $L=1501$ and the system-bath
coupling is $g=0.01$. The horizontal red dashed line denotes the steering
limit. All the parameters are dimensionless throughout this paper. The time $%
t$ was rescaled dimensionless according to the coupling constants in all the
following figures. }
\end{figure}

\section{TS inequality and TS parameter}
Let us briefly introduce a useful tool to detect steerability, i.e., the TS
parameter $S_{N}$\ which is derived from a temporal analogue of the steering
inequality~\cite{20,20-1},

\begin{equation}
S_{N}\equiv \sum\limits_{i=1}^{N}E\left( \left\langle
B_{i,t_{B}}\right\rangle _{A_{i,t_{A}}}^{2}\right) \leq 1,
\label{S_inequality}
\end{equation}%
where $N$ ($=2$, $3$) is the number of mutually-unbiased measurements that
Bob implements on his qubit and $A_{i,t_{A}}$ ($B_{i,t_{B}}$) stands for the
$i_{\text{th}}$ observable measured by Alice (Bob) at $t_{A}$ ($t_{B}$). The
violation of the above inequality indicates the existence of steerability,
otherwise, it is not sure. In the definition of steering parameter $S_{N}$,
\begin{equation}
E\left( \left\langle B_{i,t_{B}}\right\rangle _{A_{i,t_{A}}}^{2}\right)
\equiv \sum\limits_{a=\pm 1}P\left( A_{i,t_{A}}=a\right) \left\langle
B_{i,t_{B}}\right\rangle _{A_{i,t_{A}}=a}^{2},
\end{equation}%
with $P\left( A_{i,t_{A}}=a\right) $ is the probability of the Alice's
measurement to receive the result of $a=+1$ or $-1$,\ and the Bob's
expectation value conditioned on Alice's result defined as%
\begin{equation}
\left\langle B_{i,t_{B}}\right\rangle _{A_{i,t_{A}}=a}\equiv
\sum\limits_{b=\pm 1}bP\left( B_{i,t_{B}}=b|A_{i,t_{A}}=a\right) ,
\end{equation}%
where $P\left( B_{i,t_{B}}=b|A_{i,t_{A}}=a\right) $ denotes the condition
probability of Bob's measurement to get outcome $b$ (at time $t_{B}$) on the
evolved state starting from the collapsed version as Alice's measurement to
get $a$ (at time $t_{A}$).

Let us take $N=2$ for example, Alice actually has a freedom to choose her
measurement. Namely, the observables $A_{1,2}$($B_{1,2}$) can be any two of
the Pauli operators $\sigma _{x,y,z}$, more generally, can be the operators
consisting of the Pauli operators with arbitrary coefficients, as long as
the direction of $A_{1}$($B_{1}$) is perpendicular to $A_{2}$($B_{2}$).
Obviously one can expect that different measurements of Alice will produce
different steerability, and which will be shown in the following sections.
When associating TS with the error rate of a channel in BB84 protocol, the
steering inequality bound and the BB84 threshold are found to be equivalent~%
\cite{20}. Namely, when $S_{2}$ [in Eq.\thinspace (\ref{S_inequality}) for $%
N=2$] becomes larger than $1$, both of the steerability and security of the
channel will be verified. For a unitary evolution, there will always be $%
S_{2}>1$ in the whole duration, whereas, the quantum decoherence will
significantly shorten the security time intervals~\cite{20}.

\textbf{TS parameter in a two-dimensional dephasing channel. }Now let us
calculate the TS parameter based on an arbitrary choice of measurement
pairs. By using two kinds of rotation: $U_{x,\theta }\equiv e^{i\frac{\theta
}{2}\sigma _{x}}$, and $U_{y,\phi }\equiv e^{i\frac{\phi }{2}\sigma _{y}}$,
the measurement bases can be described as $\left\vert \psi
_{1}^{(+)}\right\rangle =U_{y,\phi }U_{x,\theta }\left\vert 1\right\rangle $%
, $\left\vert \psi _{1}^{(-)}\right\rangle =U_{y,\phi }U_{x,\theta +\pi
}\left\vert 1\right\rangle $, $\left\vert \psi _{2}^{(+)}\right\rangle
=U_{y,\phi }U_{x,\theta +\pi /2}\left\vert 1\right\rangle $, and $\left\vert
\psi _{2}^{(-)}\right\rangle =U_{y,\phi }U_{x,\theta +3\pi /2}\left\vert
1\right\rangle $. Note that $\left\vert \psi _{1}^{(\pm )}\right\rangle $
are the eigenstates of the observable of $A_{1}$ and $\left\vert \psi
_{2}^{(\pm )}\right\rangle $ are responsible for $A_{2}$. If we add a
rotation around $\hat{z}$ axis to $\left\vert \psi _{2}^{(+)}\right\rangle $%
, it will turn to other directions on the plane perpendicular to $A_{1}$.
Whereas, all the rotation around $\hat{z}$ direction causes no change to the
TS parameter. By varying $\theta $ and $\phi $, one can reach the
eigenstates corresponding to an arbitrary observable pair $\{A_{1},A_{2}\}$.
When choosing the initial state of the system as a maximally mixed state,
i.e., $\rho _{s}\left( 0\right) =\mathbb{I}/2$, we obtain a general form of
the TS parameter that%
\begin{equation}
S_{2}=E_{1}+E_{2},  \label{S2}
\end{equation}%
with the quantity:
\begin{widetext}
\begin{eqnarray}
E_{1} &=&P_{1}^{(+)}\left\{ \text{Tr}\left[ \left\vert \psi
_{1}^{(+)}\right\rangle \left\langle \psi _{1}^{(+)}\right\vert \rho
_{1t}^{(+)}\right] -\text{Tr}\left[ \left\vert \psi _{1}^{(-)}\right\rangle
\left\langle \psi _{1}^{(-)}\right\vert \rho _{1t}^{(+)}\right] \right\} ^{2}
\notag \\
&&+P_{1}^{(-)}\left\{ \text{Tr}\left[ \left\vert \psi
_{1}^{(+)}\right\rangle \left\langle \psi _{1}^{(+)}\right\vert \rho
_{1t}^{(-)}\right] -\text{Tr}\left[ \left\vert \psi _{1}^{(-)}\right\rangle
\left\langle \psi _{1}^{(-)}\right\vert \rho _{1t}^{(-)}\right] \right\}
^{2},\notag\\
\end{eqnarray}
\end{widetext}
where $P_{1}^{(\pm )}=$Tr$\left[ \left\vert \psi _{1}^{(\pm
)}\right\rangle \left\langle \psi _{1}^{(\pm )}\right\vert \rho _{s}\left(
0\right) \right] $ denotes the probability of Alice's measurement on the
initial state of the system\ and $\rho _{1t}^{(\pm )}$ displays the evolved
state from $\left\vert \psi _{1}^{(\pm )}\right\rangle \left\langle \psi
_{1}^{(\pm )}\right\vert $ under the quantum channel, consequently, we have

\begin{equation}
E_{1}=\left[ \cos ^{2}\theta \cos ^{2}\phi \left( 1-\text{Re}F_{t}\right) +%
\text{Re}F_{t}\right] ^{2},
\end{equation}%
with $F_{t}$ corresponding to the decoherence factor causing by the
dephasing channel in Eq.\thinspace (\ref{rho_t}). Similarly, over the state $%
\left\vert \psi _{2}^{(\pm )}\right\rangle $, one can obtain
\begin{equation}
E_{2}=\left[ \sin ^{2}\theta \cos ^{2}\phi \left( 1-\text{Re}F_{t}\right) +%
\text{Re}F_{t}\right] ^{2}.
\end{equation}%
We shall emphasize that the above expressions generally demonstrate the
dependence of the TS parameter $S_{2}$ on the decoherence factor in a
dephasing channel. Different values of $\theta $ and $\phi $ reflect the
contribution of different measurement choices on $S_{2}$. Obviously, $S_{2}$
can reach its maximal value\ when $\theta $ and $\phi $ are set to satisfy $%
\sin 2\theta =0$ and $\sin \phi =0$, namely, $A_{1}=\sigma _{z}$ and $%
A_{2}=\sigma _{y}$ (or $\sigma _{x}$), and thus
\begin{equation}
S_{2,\max }=1+\text{Re}^{2}F_{t}.
\end{equation}%
We can also consider the case of three measurements operated by Alice, then
another observable $A_{3}$ is introduced, whose direction is perpendicular
to both $A_{1}$ and $A_{2}$ and the corresponding eigenstates can be
described as $\left\vert \psi _{3}^{(\pm )}\right\rangle =U_{y,\phi
}\left\vert 1\right\rangle $. Then we have the third quantity
\begin{equation}
E_{3}=\left( \cos ^{2}\phi \text{Re}F_{t}+\sin ^{2}\phi \right) ^{2},
\end{equation}%
and thus the TS parameter becomes
\begin{equation}
S_{3}=E_{1}+E_{2}+E_{3},
\end{equation}%
For the case of $A_{1}=\sigma _{z}$, $A_{2}=\sigma _{y}$,\ and $A_{3}=\sigma
_{x}$, we have the maximum
\begin{equation}
S_{3,\max }=1+2\text{Re}^{2}F_{t}.
\end{equation}%
Obviously, the maximal values of $S_{2,\max }$ and $S_{3,\max }$ are not
less than $1$ and their dynamics are absolutely dominated by the real part
of the decoherence factor, i.e., Re$F_{t}$, which implies that before Re$%
F_{t}$ decays to zero, proper observable pair of Alice such as $\left\{
\sigma _{z},\sigma _{x}\right\} $ (or $\left\{ \sigma _{z},\sigma
_{y}\right\} $) make sure the existence of the bidirectional-measurement TS.
Furthermore, a choice of the measurements $\left\{ \sigma _{z},\sigma
_{x},\sigma _{y}\right\} $ guarantees the persistence of the
tridirectional-measurement TS. However, after a sufficiently long-time
decoherence process, $S_{2,\max }$ and $S_{3,\max }$ will tend to $1$, then
the existence of TS becomes not sure.

In Fig.\thinspace 1, we plot TS parameter $S_{2}$ vs evolution time with a
duration $t_{b}=5$. The parameter $\gamma =1$. The coupling parameter is
chosen as $g=0.01$ throughout the following numerical calculations. The
measurement bases are chosen as $\left\vert +\right\rangle $ ($\left\vert
-\right\rangle $) and $\left\vert R\right\rangle $ ($\left\vert
L\right\rangle $) which are the eigenstates of $\sigma _{x}$ and $\sigma
_{y} $. This choice of measurement bases leads to a non-maximal $S_{2}$ and
it behaves as an oscillation with time. We investigate the cases of $\lambda
=0.2$, $1$ and $1.2$ in subfigures (a), (b) and (c), respectively. When the
environment undergoes the criticality at $\lambda =1$ [see subfigure (b)],
the peak values of TS are obviously suppressed below the steering limit ($%
S_{N,\text{Lim}}=1$ denoted by red dashed lines) after a certain time.
Whereas, beyond the critical region [see subfigures (a) and (c)], $S_{2}$
periodically crosses the steering limit and the peak values keep higher than
$S_{N,\text{Lim}}=1$. The above comparison of the dynamical behavior of $%
S_{2}$ reveals the significant influence of the quantum criticality of the
environment on the TS property of the coupled system.
\begin{figure}[tbp]
\begin{center}
\includegraphics[width=0.5\textwidth, height=0.37\textwidth]{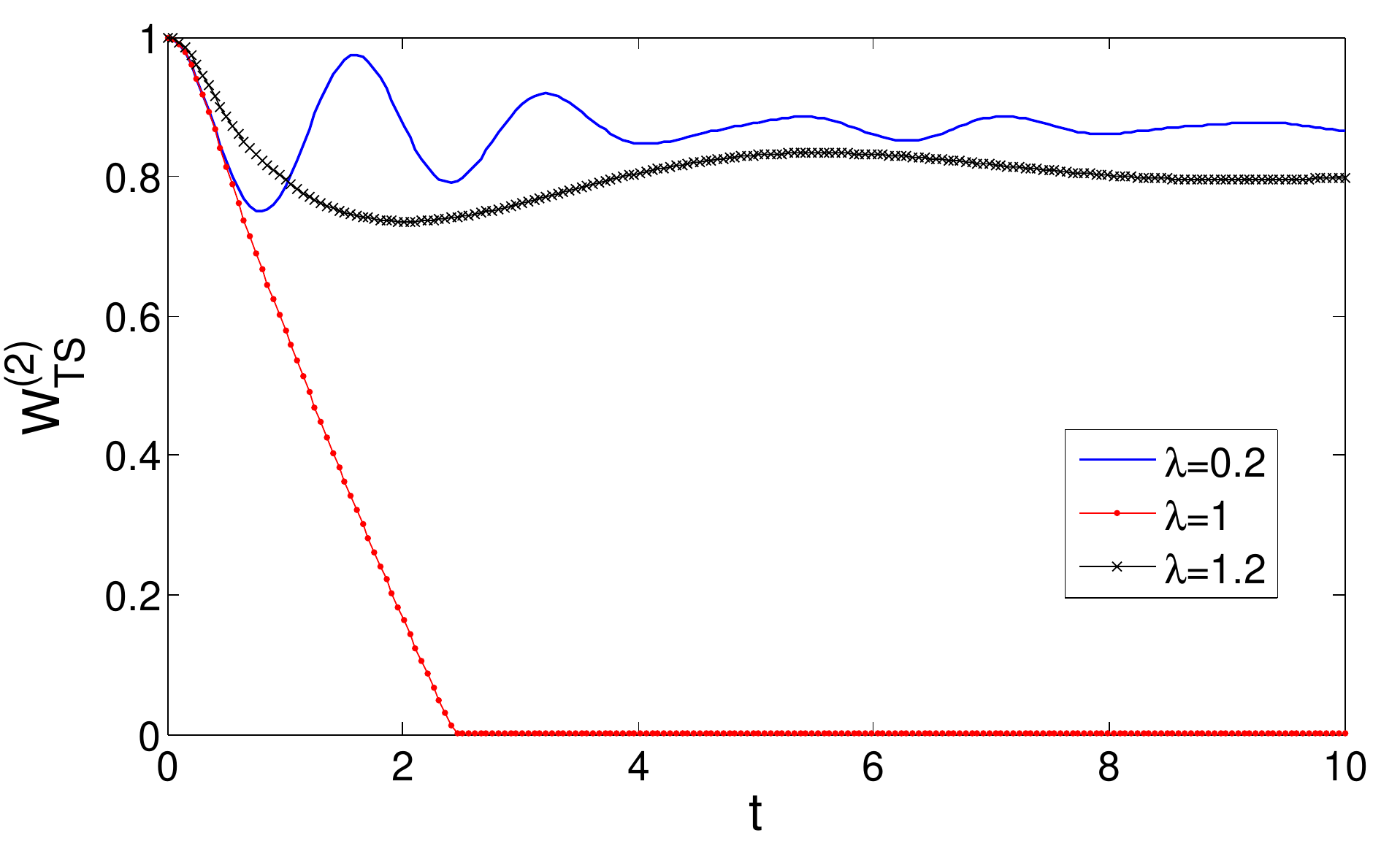}
\end{center}
\caption{
(Color online) Bidirectional-measurement TS weight versus time $t$
for different parameters $\protect\lambda =0.2,~1$ and $1.2$. The parameter $%
\protect\gamma =1$, the size of the surrounding spin chain is $L=1501$ and
the system-bath coupling is $g=0.01$. }
\end{figure}

\begin{figure}[tbp]
\begin{center}
\includegraphics[width=0.5\textwidth, height=0.37\textwidth]{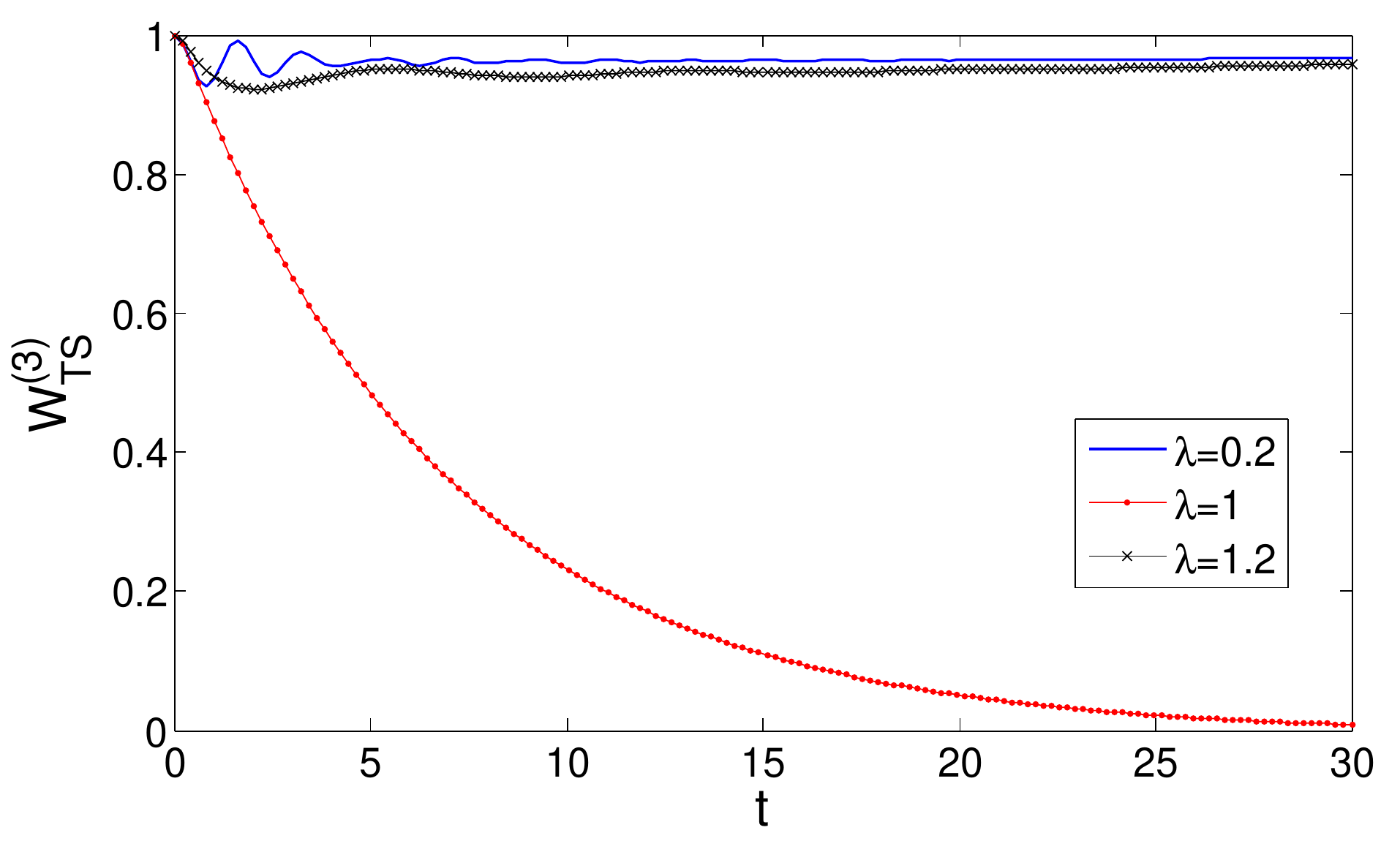}
\end{center}
\caption{
(Color online) Tridirectional-measurement TS weight versus time $t$
for different parameters $\protect\lambda =0.2,~1$ and $1.2$. The parameter $%
\protect\gamma =1$, the size of the surrounding spin chain is $L=1501$ and
the system-bath coupling is $g=0.01$. }
\end{figure}

\section{TS weight and its power}
\textbf{TS weight.} In order to precisely quantify the TS, a direct analogue
of EPR steerable weight named TS weight is introduced via semidefinite
programing as~\cite{21}
\begin{equation}
W_{\text{TS}}^{(N)}\equiv 1-\max \text{ Tr}\sum\limits_{\lambda }^{m^{N}}%
\tilde{\sigma}_{\lambda },  \label{TSW_1}
\end{equation}%
subject to
\begin{equation}
\sigma _{a|A_{i}}-\sum\limits_{\lambda }D_{\lambda }\left( a|A_{i}\right)
\tilde{\sigma}_{\lambda }\geq 0,\text{ \ }\forall a,A_{i},  \label{TSW_2}
\end{equation}%
and
\begin{equation}
\tilde{\sigma}_{\lambda }\geq 0,
\end{equation}%
where $\{\sigma _{a|A_{i}}\}$ stands for the assemblage received by Bob and
whose origin is the resulting state after Alice's measurements and a
following evolution under a quantum channel. $N$ denotes the number of the
observables measured by Alice and a $m$-dimensional operator $A_{i}$ ($%
i=1...N$)\ represents the $i_{\text{th}}$ observable chosen by Alice and $a$
is the measurement result. Then Bob only cares whether the assemblage he
receives can be written in a hidden-state form~\cite{16,21}, e.g., the
second term in Eq.\thinspace (\ref{TSW_2}), where $\lambda $ indicates a
local hidden variable, correspondingly, a set of semidefinite matrices $%
\tilde{\sigma}_{\lambda }$ act as the basis matrices, held by Bob,
cooperating with the extremal deterministic single party conditional
probability $D_{\lambda }\left( a|A_{i}\right) $. If there does not exists
steerability between Alice's and Bob's system, at least one of the
hidden-state form can be found to classically fabricate Alice's measurement
result. And thus $W_{\text{TS}}^{(N)}$ in Eq.\thinspace (\ref{TSW_1})
outputs a zero value. Otherwise, the steerability is sufficiently and
necessarily verified.
\begin{figure}[tbp]
\begin{center}
\includegraphics[width=0.5\textwidth, height=0.37\textwidth]{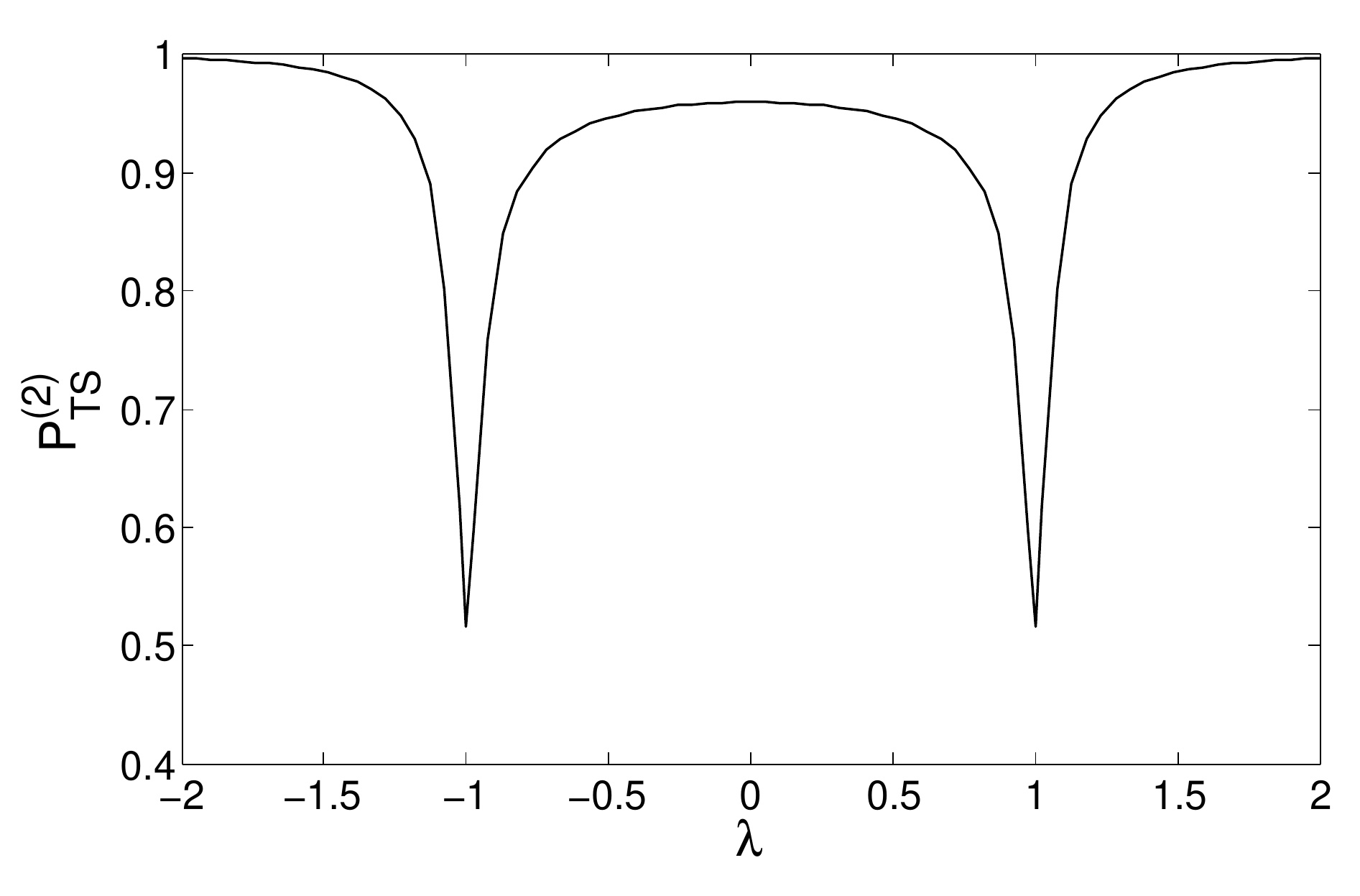}
\end{center}
\caption{
(Color online) Bidirectional-measurement TS weight power versus
parameter $\protect\lambda $ for $\protect\gamma =1$. The evolution duration
is bounded by $t_{b}=10$. The size of the surrounding spin chain is $L=1501$
and the system-bath coupling is $g=0.01$. }
\end{figure}

\begin{figure}[tbp]
\begin{center}
\includegraphics[width=0.5\textwidth, height=0.45\textwidth]{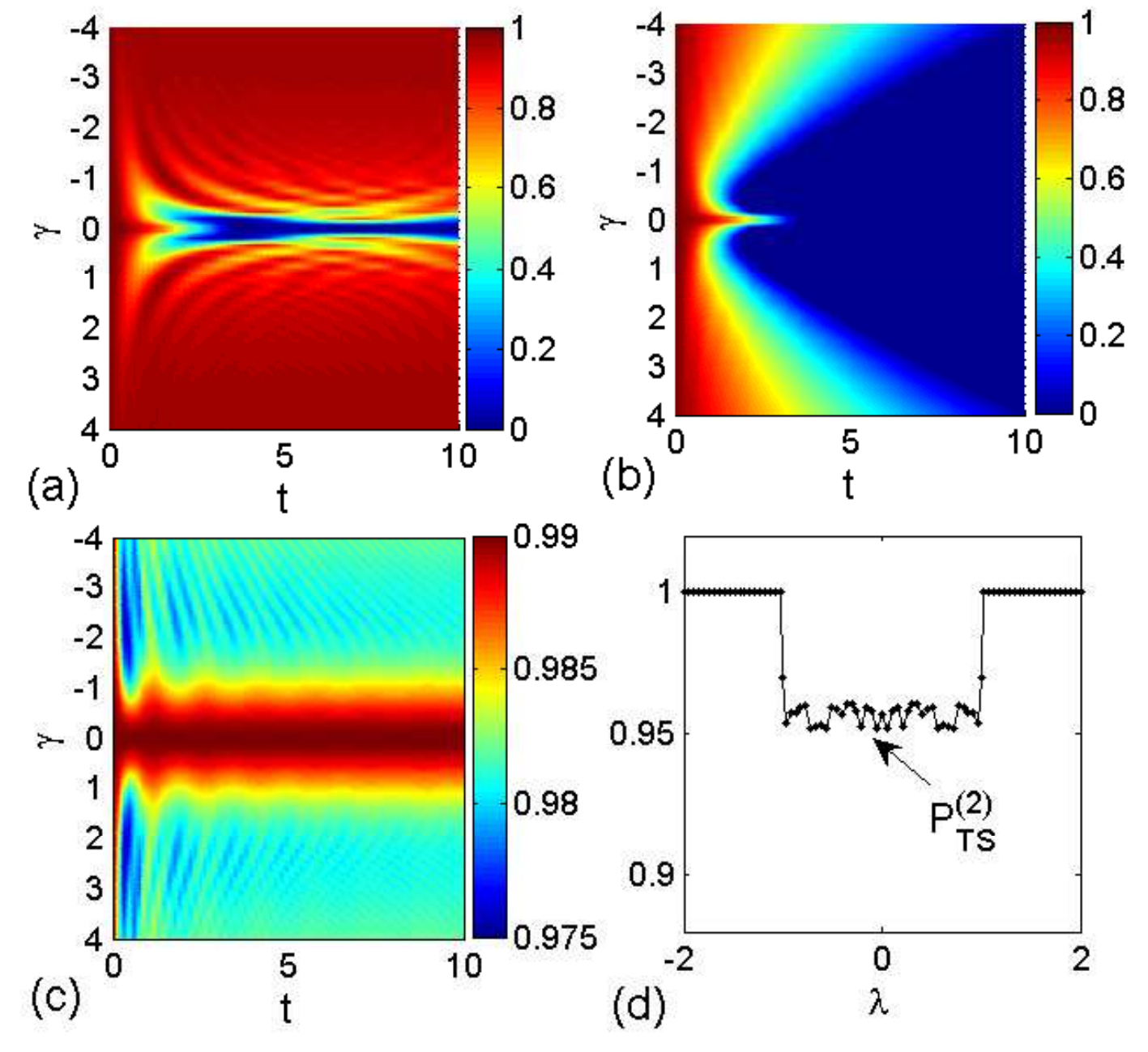}
\end{center}
\caption{
(Color online) Bidirectional-measurement TS weights versus both
time $t$ and $\protect\gamma $ are shown by the contours in subfigures (a),
(b) and (c) corresponding to the cases $\protect\lambda =0.2,~1$ and $1.2$,
respectively. The size of the surrounding spin chain is $L=1501$ and the
system-bath coupling is $g=0.01$. (d) presents the bidirectional-measurement
TS weight power versus different $\protect\lambda $ for the case of $\protect%
\gamma =0.002$ (close to zero). }
\end{figure}
We numerically study a bidirectional-measurement TS weight $W_{\text{TS}%
}^{(2)}$\ versus time $t$ for different field strengths of $\lambda =0.2$, $%
1 $, and $1.2$ and show the results in Fig.\thinspace 2. Our results clearly
show that the quantum criticality in the environment seriously destroys the
TS.\ When the parameter $\lambda $ is adjusted to the critical point $%
\lambda _{c}=1$, the TS weight $W_{\text{TS}}^{(2)}$ decays rapidly and
displays a \textquotedblleft sudden death\textquotedblright . This numerical
result agrees well with the heuristic analysis of the decoherence factor in
Eq.\thinspace (\ref{Ft-smallk}). Other cases beyond the critical region,
such as $\lambda =0.2$ and $1.2$, $W_{\text{TS}}^{(2)}$ oscillates slightly
around a high value about $0.9$ and $0.8$, respectively. Recalling the
results in Fig.\thinspace 1, after the time about $t>2.4$, the parameter $%
S_{2}$ becomes lower than $1$, however, it does not make sure whether there
exists TS or not. Fortunately, the precise quantity $W_{\text{TS}}^{(2)}$
verifies the vanishment of TS.

In order to show the influence of different choices of measurement on the
TS, in Fig.\thinspace 3, we consider the tridirectional-measurement-induced
TS characterized by $W_{\text{TS}}^{(3)}$. Differently, the quantum
criticality suppresses $W_{\text{TS}}^{(3)}$ to zero asymptotically unlike
the \textquotedblleft sudden death\textquotedblright\ of $W_{\text{TS}%
}^{(2)} $ (in Fig.\thinspace 2). Moreover, the decay rate is much more
slowly than that of $W_{\text{TS}}^{(2)}$ (in Fig.\thinspace 2). For the
cases of $\lambda =0.2$ and $1.2$, the average values of $W_{\text{TS}%
}^{(3)} $ are obviously higher than these of $W_{\text{TS}}^{(2)}$ (in
Fig.\thinspace 2).
Doing more measurements is generally helpful to capture more information of the target state, and thus it is expected to be beneficial for quantum steering. In the reference~\cite{16}, the authors numerically show the fact that increasing measurement direction numbers will increase quantum steering weight. Therefore, when the quantum decoherence is considered, the steering proposal with more measurement directions will be more robust to decoherence. Consequently, our results show that the steerability with two-directional measurements vanishes after a certain time (red line in Fig.~2). However, in the case of three-directional measurements, the steerability will keep non-zero values for quite a long time (red line in Fig.~3). We note that the sudden vanishing phenomena were found in several kinds of nonclassical effects~\cite{Bartkowiak,Miranowicz,LiuYX,WangXG,Lambert,MaJ}.

On the other hand, from the analytical result of the decoherence factor in Eq.~(\ref{Ft}) and the approximate result in Eq.~(\ref{Ft-smallk}), one can find that only near the critical point, there is a monotonic decay of the quantum coherence. Other than that, the quantum coherence dynamics behaves like periodic function of time, which can be understood as the information periodically flow back to the system from the bath. Because of the close relationship between quantum coherence and steerability, the dramatic decoherence at the critical point leads to much lower steerability (red lines in Fig.~2 and Fig.~3) than those cases beyond the critical region (blue and black lines in Fig.~2 and Fig.~3).

\textbf{TS weight power.} In order to highlight the power of a quantum
channel in influencing TS, we introduce a concept of \textit{TS weight power
}$P_{\text{TSW}}^{(N)}$ as%
\begin{equation}
P_{\text{TSW}}^{(N)}\equiv \underset{\{A_{i}\}}{\max }\frac{1}{T_{b}}%
\int_{0}^{T_{b}}W_{\text{TS}}^{(N)}(t)\text{d}t,  \label{TSWP_1}
\end{equation}%
where the maximalization is obtained over all the possible observables $A_{i}
$ in an assemblage consists of $N$-directional\ measurements, then a time
average is performed over an\ evolution duration $T_{b}$. In the definition
of $P_{\text{TSW}}^{(N)}$, the influences of Alice's choice of measurements
and the temporal evolution on the steerability are technically covered,
therefore, the role of the quantum channel becomes prominent.

In Fig.\thinspace 4, we plot $P_{\text{TSW}}^{(2)}$ vs the coupling strength
parameter $\lambda $ for the case of $\gamma =1$. By randomly choosing a
sufficient number of observables along two directions, we numerically
complete the maximization process in the definition of Eq.\thinspace (\ref%
{TSWP_1}). From the results, the two minimum points precisely indicate the
critical points of the environment.

When we take into account another values of $\gamma $, in Fig.\thinspace 5,
we plot $W_{\text{TS}}^{(2)}$ versus parameter $\gamma $ and time $t$ in
subfigures (a), (b) and (c) corresponding to $\lambda =0.2$, $1$ and $1.2$,
respectively. From the contour maps, one can clearly see the critical region
drawn by $\gamma \rightarrow 0$. For a small parameter such as $\lambda =0.2$%
, the sudden death of TS appears only near the critical region $\gamma =0$.
While, for the case $\lambda =1$ (a critical value), the phenomenon of the
sudden death of TS exists for all the range of $\gamma $ as long as the
evolution is long enough. This is consistent with the fact that for any
values of $\gamma $, QPTs can occur at the critical point of $\lambda _{c}=1$%
. However, larger strength such as $\lambda =1.2$ prohibits the sudden death
of TS for any values of $\gamma $. Instead, $W_{\text{TS}}^{(2)}$ displays
weak oscillations around the values close to $1$. In order to clearly show
the critical region, we calculate the power $P_{\text{TSW}}^{(2)}$ versus
different $\lambda $ in the case of $\gamma =0.002$ (close to zero). When
the environment undergoes the quantum critical region bounded by $\left\vert
\lambda \right\vert \leq 1$, $P_{\text{TSW}}^{(2)}$ enters a disorder region
which sensitively indicates the critical region.

\section{Conclusion}

In this paper, we have investigated the TS of a qubit surrounded by a
\textit{XY }spin chain where rich QPTs take place. The system-bath
interaction has been assumed in a dephasing form, and the decoherence factor
was obtained analytically. Based on this dephasing channel, we have given
the analytical expression of the TS parameter which is directly dependent on
the decoherence factor. Numerical results showed that at the critical point $%
\lambda =1$, the values of the TS parameter was obviously suppressed below
the steering limit after a certain time. The disappearance of TS was
confirmed when we took into account the TS weight. The sudden death of TS
was found when the environment underwent its QPT. Finally, we have developed
a new concept named \textit{TS weight power }to quantify the capacity of the
channel in influencing TS. With its help, the criticality of the environment
can be clearly indicated. All of our analytical and numerical results
reflect the significant influence of the quantum criticality on the TS
dynamics. From another perspective, our results also reveal that the TS in
such an auxiliary qubit can play as an effective tool to detect the QPTs of
the surroundings. It still remains future interests of investigation on
other quantum correlation dynamics in a wide variety of quantum-crical
environments.

\section*{Appendix: \textbf{Heuristic analysis of the decoherence factor}}

\textbf{\ }We calculate the norm of the decoherence factor,
\begin{eqnarray}
\left\vert F_{t}\right\vert  &=&\prod_{k>0}\left\vert F_{t}^{k}\right\vert
\notag \\
&=&\prod_{k>0}\big \{1-\big[\cos \left( t\Omega _{k,-}\right) \sin \left(
t\Omega _{k,+}\right) \sin \alpha _{k,+}  \notag \\
&&-\sin \left( t\Omega _{k,-}\right) \cos \left( t\Omega _{k,+}\right) \sin
\alpha _{k,-}\big]^{2}  \notag \\
&&-\sin ^{2}\left( t\Omega _{k,+}\right) \sin ^{2}\left( t\Omega
_{k,-}\right) \sin ^{2}\left( \alpha _{k,+}-\alpha _{k,-}\right) \big \}^{%
\frac{1}{2}},  \notag \\
&&  \label{absF_mn}
\end{eqnarray}%
where the parameters $\Omega _{k,\pm }$ and $\alpha _{k,\pm }=\theta
_{k}-\theta _{k,\pm }$ can be obtained from Eq.\thinspace (\ref{p2}) by
replacing $\lambda $ with ${\lambda }_{\pm }={\lambda \pm g}$. Obviously,
the values of $\left\vert F_{t}^{k}\right\vert $ are less than unit.
Therefore, in the large $L$\ limit, $\left\vert F_{t}\right\vert $ is
expected to decay to zero.

By similar analysis of Ref.~\cite{quan,Zhe07}, we introduce a cutoff number $%
K_{c}$\ to provide a lower bound of the norm of the decoherence factor as
\begin{equation}
\left\vert F_{t}\right\vert _{K_{c}}=\prod_{k>0}^{K_{c}}\left\vert
F_{t}^{k}\right\vert \geq \left\vert F_{t}\right\vert ,  \label{F_Kc}
\end{equation}%
therefore, we have the logarithmic form
\begin{equation}
S\left( t\right) =\ln \left\vert F_{t}\right\vert _{K_{c}}\equiv
-\sum_{k>0}^{K_{c}}\left\vert \ln F_{t}^{k}\right\vert .
\end{equation}%
Now let us consider some conditions that $k$ is small enough, $L$ is very
large, and $\gamma $ is a finite number, then we could omit some small terms
in higher orders. Thus we have the following approximations:\ $\Omega
_{k,\pm }\approx -2\left\vert 1-{\lambda }_{\pm }\right\vert $, and $\sin
\alpha _{k,\pm }\approx \pm 2g\gamma \pi k/\left( L\left\vert 1-\lambda
\right\vert \left\vert 1-{\lambda }_{+}\right\vert \right) $ consequently $%
\sin \left( \alpha _{k,+}-\alpha _{k,-}\right) \approx 4g\gamma \pi k/\left(
L\left\vert \left( 1-{\lambda }_{+}\right) \left( 1-\lambda _{-}\right)
\right\vert \right) $. Furthermore, the approximated value of $S\left(
t\right) $ reads
\begin{eqnarray}
S\left( t\right) &\approx &-E\left( K_{c}\right) \left( 1-{\lambda }%
_{+}\right) ^{-2}\left( 1-{\lambda }_{-}\right) ^{-2}\left\vert 1-\lambda
\right\vert ^{-2}\gamma ^{2}g^{2}  \notag \\
&&\times \big\{4\sin ^{2}\left( 2t\left\vert 1-\lambda _{+}\right\vert
\right) \sin ^{2}\left( 2t\left\vert 1-\lambda _{-}\right\vert \right)
\left\vert 1-\lambda \right\vert ^{2}  \notag \\
&&+[\cos \left( 2t\left\vert 1-\lambda _{-}\right\vert \right) \sin \left(
2t\left\vert 1-\lambda _{+}\right\vert \right) \left\vert 1-\lambda
_{-}\right\vert  \notag \\
&&+\sin \left( 2t\left\vert 1-\lambda _{-}\right\vert \right) \cos \left(
2t\left\vert 1-\lambda _{+}\right\vert \right) \left\vert 1-\lambda
_{+}\right\vert ]^{2}\big\},  \notag \\
&&  \label{S}
\end{eqnarray}%
where $E\left( K_{c}\right) =2\pi ^{2}K_{c}\left( K_{c}+1\right) \left(
2K_{c}+1\right) /\left( 6L^{2}\right) .$ Now let us consider a short time
and a weak enough coupling $g$, when $\lambda $ approaches to $\lambda
_{c}=1,$ there will be $S_{c}\left( t\right) \approx -16E\left( K_{c}\right)
\gamma ^{2}g^{2}\left[ 4t^{4}+t^{2}\left( 1-\lambda \right) ^{-2}\right] $.

Note that when $L\rightarrow \infty $ in thermodynamic limit, the parameter$%
\ E\left( K_{c}\right) $ should tend to $0$. In this case, due to the
contribution of $\left( 1-\lambda \right) ^{-2}$, the terms with $t^{2}$
survive. Thus, the norm of the factor approximately behaves as
\begin{equation}
\left\vert F_{t}\right\vert _{K_{c}}\approx e^{-R_{c}t^{2}},
\end{equation}%
with a decay rate $R_{c}=16E\left( K_{c}\right) \gamma ^{2}g^{2}\left(
1-\lambda \right) ^{-2}.$ In some cases that the factor $E\left(
K_{c}\right) \gamma ^{2}g^{2}$ approaches to$\ 0$, the decay of the $%
\left\vert F\left( t\right) \right\vert $ is still possible when the
parameter enter the critical region, i.e., $\lambda \rightarrow 1$ due to
the denominator $\left( 1-\lambda \right) ^{2}$. However, beyond the
critical region, the decay of the bound $\left\vert F_{t}\right\vert
_{K_{c}} $ will not be sure. Instead, it perhaps behave as an oscillation
with time due to the periodic functions such as the $\sin \left(
2t\left\vert 1-\lambda _{+}\right\vert \right) $ in Eq.\thinspace (\ref{S}).
The analysis above reveals the fact that the remarkable decay of $\left\vert
F\left( t\right) \right\vert $ corresponds to the occurrence of\ quantum
criticality at the critical point $\lambda _{c}=1$ for any values of $\gamma
$.

\section{Acknowledgments}
Z.S. is supported by the National Natural Science Foundation of China under
Grant No. 11375003 and 11775065, the Program for HNUEYT under Grant No.
2011-01-011, the Zhejiang Natural Science Foundation with Grant Nos.
LZ13A040002 and LY17A050003, the Ministry of Science and Technology of China
under Grant No. 2016YFA0301802, the funds for the Hangzhou-City Quantum
Information and Quantum Optics Innovation Research Team. B.L. acknowledges
the Thirteenth Five-year Planning Project of Jilin Provincial Education
Department Foundation under Grant No.JJKH20170650KJ. Y.H. acknowledges
support by the Natural Science Foundation of Zhejiang province of China NO.
LQ16A040001 and NSFC through Grant No. 11605157.

%\bibliography{citebook}
%\input{main.bbl}

\end{document}